\documentclass[journal]{IEEEtran}

\usepackage{epsfig,amsmath,amssymb,epsf,amsthm,scalefnt,multirow,subfig}
\usepackage{xcolor}
\usepackage{float}
\usepackage{cite}
\usepackage{psfrag}
\usepackage{gensymb}
\usepackage{cleveref}
\usepackage{mathrsfs}
\usepackage{mathtools}
\usepackage{algpseudocode}
\usepackage{algorithm}
\usepackage{enumerate}

\newcommand{\argmax}{\mathop{\mathrm{argmax}}}
\newcommand{\argmin}{\mathop{\mathrm{argmin}}}

\newtheorem*{lemma*}{Lemma}

  \def\cC{{\mathcal{C}}}

 \def\cN{{\mathcal{N}}}  
\def\cQ{{\mathcal{Q}}}

\def\argmin{\mathop{\mathrm{argmin}}}
\def\argmax{\mathop{\mathrm{argmax}}}
\def\diag{\mathop{\mathrm{diag}}}

\def\b0{{\pmb{0}}} 

\def\ba{{\mathbf{a}}}   
 \def\bp{{\mathbf{f}}}  \def\bh{{\mathbf{h}}}
   
 \def\bn{{\mathbf{n}}}  \def\bp{{\mathbf{p}}}
   
 \def\bv{{\mathbf{v}}} \def\bw{{\mathbf{w}}} 
\def\by{{\mathbf{y}}}   

\def\bA{{\mathbf{A}}}  \def\bC{{\mathbf{C}}} \def\bD{{\mathbf{D}}}
 \def\bF{{\mathbf{F}}}  \def\bH{{\mathbf{H}}}
\def\bI{{\mathbf{I}}}   
 \def\bN{{\mathbf{N}}} \def\bO{{\mathbf{O}}} \def\bP{{\mathbf{P}}}
   \def\bT{{\mathbf{T}}}
\def\bU{{\mathbf{U}}} \def\bV{{\mathbf{V}}}  \def\bX{{\mathbf{X}}}

\begin{document}

\title{Downlink Channel Reconstruction \\ for Spatial Multiplexing in Massive MIMO Systems}

\author{Hyeongtaek~Lee,
        Hyuckjin~Choi,
        Hwanjin~Kim,
        Sucheol~Kim,
  		Chulhee~Jang,
  		Yongyun~Choi,
        and~Junil~Choi
%        ~\IEEEmembership{Member,~IEEE.}% <-this % stops a space

\thanks{H. Lee, H. Choi, H. Kim, S. Kim and J. Choi are with the School of Electrical Engineering, Korea Advanced Institute of Science and Technology (e-mail: htlee8459@kaist.ac.kr;  hugzin008@kaist.ac.kr; jin0903@kaist.ac.kr; loehcusmik@kaist.ac.kr; junil@kaist.ac.kr).}

\thanks{C. Jang and Y. Choi are with the Network Business, Samsung Electronics Co., LTD (e-mail: chulhee.jang@samsung.com; yongyun.choi@samsung.com).}
}

%\thanks{M. Shell was with the Department
%of Electrical and Computer Engineering, Georgia Institute of Technology, Atlanta,


\maketitle

\begin{abstract}
To get channel state information (CSI) at a base station (BS), most of researches on massive multiple-input multiple-output (MIMO) systems consider time division duplexing (TDD) to get benefit from the uplink and downlink channel reciprocity. Even in TDD, however, the BS still needs to transmit downlink training signals, which are referred to as channel state information reference signals (CSI-RSs) in the 3GPP standard, to support spatial multiplexing in practice. This is because there are many cases that the number of transmit antennas is less than the number of receive antennas at a user equipment (UE) due to power consumption and circuit complexity issues. Because of this mismatch, uplink sounding reference signals (SRSs) from the UE are not enough for the BS to obtain full downlink MIMO CSI. Therefore, after receiving the downlink CSI-RSs, the UE needs to feed back quantized CSI to the BS using a pre-defined codebook to support spatial multiplexing. In this paper, possible approaches to reconstruct full downlink MIMO CSI at the BS are proposed by exploiting both the SRS and quantized downlink CSI considering practical antenna structures with reduced downlink CSI-RS overhead. Numerical results show that the spectral efficiencies by spatial multiplexing based on the proposed downlink MIMO CSI reconstruction techniques outperform the conventional methods solely based on the quantized CSI.
\end{abstract}
\section{Introduction} \label{Introduction}

\IEEEPARstart{M}{assive} multiple-input multiple-output (MIMO) systems, which deploy tens or hundreds of antennas at a base station (BS), have become one of the key features of future wireless communication systems including the upcoming fifth generation (5G) cellular networks \cite{Rusek:2013,Bjornson:2016mag,Larsson:2014,Marzetta:2010,Hoydis:2013}. It is now well known that massive MIMO systems can effectively mitigate inter-user interference with simple linear precoders (for downlink) and receive combiners (for uplink) and achieve high spectral efficiency by supporting a large number of users simultaneously \cite{Hoydis:2013,Marzetta:2010}. 

All the benefits mentioned above are possible only when the BS has accurate channel state information (CSI). Although frequency division duplexing (FDD) dominates current wireless communication systems, FDD massive MIMO suffers from excessive downlink training and uplink feedback overheads \cite{Love:2008,Choi:closedloop_FDD,Larsson:2014,Marzetta:2010,Hoydis:2013}. There has been much work on resolving these issues \cite{Gao:2016(FDDoverhead),Fang:2017(FDDoverhead),Shen:2015(FDDoverhead),Shen:2018(FDDoverhead),Rao:2014(FDDoverhead),Han:2019,Gao:2015,Hyuckjin:2020(FDDoverhead)}. Especially, \cite{Han:2019,Gao:2015,Hyuckjin:2020(FDDoverhead)} exploited spatial reciprocity between the downlink and uplink channels in FDD. Since both channels experience the same environment and share some dominant channel parameters, e.g., path delays and directions, even in FDD, the proposed approaches could remove most of the overheads in FDD massive MIMO. These works are, however, restricted to a single antenna user equipment (UE) case,~and it is not straightforward to extend the techniques to multiple~antennas at the UE side. When the UE has multiple antennas, \cite{Rao:2014(FDDoverhead),Alkhateeb:2014,Vlachos:2019} proposed to exploit the sparse nature of channels and compressive sensing techniques to mitigate the channel training overhead. Since not all channels would experience the sparsity, however, it is difficult to extend these approaches into more general environments. 

The most common and direct approach to get rid of all downlink training and uplink feedback issues is to adopt time division duplexing (TDD) to exploit the downlink and uplink channel reciprocity
\cite{Marzetta:2010,Guey:2004,Jose:2011,Rusek:2013,Hoydis:2013,Bjornson:2014,Hoydis:2013Bell,Ngo:2013,Bjornson:2017,Ngo:2017TWC,Mishra:2019}. In TDD, exploiting the channel reciprocity is useful especially when the BS supports multiple users simultaneously with a single data stream per UE through multi-user MIMO, which have been the main focus of most of massive MIMO researches. However, single-user (SU) MIMO with spatial multiplexing, which has been neglected from massive MIMO researches so far, is still important in practice and must be optimized for massive MIMO as well.

For spatial multiplexing at the BS, however, exploiting the downlink and uplink channel reciprocity may be insufficient even in TDD. It is common in practice, but has not been taken into consideration in most of MIMO researches before, that the UE is deployed with the number of transmit antennas less than the number of receive antennas because of many practical constraints including power consumption and circuit complexity issues at the UE side \cite{G.Liu:2018,Y.Liu:2019}. Therefore, the uplink sounding reference signals (SRSs), which are only transmitted from the transmit antennas at the UE, are not enough for the BS to obtain full downlink MIMO CSI by exploiting the channel reciprocity.\footnote{This problem also hinders from exploiting the proposed techniques in \cite{Han:2019,Gao:2015,Hyuckjin:2020(FDDoverhead)} when the UE has multiple antennas since the number of transmit antennas is less than that of receive antennas.} This is why the 3GPP standard defines downlink channel training using channel state information reference signals (CSI-RSs) and CSI quantization codebooks (or precoding matrix indicator (PMI) codebooks) even for TDD \cite{3GPP:38.214}.  
	
It is critical to reduce the downlink CSI-RS overhead for massive MIMO, and most effective way to reduce the overhead is by grouping multiple antennas at the BS as a single antenna port \cite{AntGrouping1,AntGrouping2}. Each antenna port transmits the same CSI-RS while the antennas in a port can have different weights to beamform the CSI-RS \cite{AntGrouping3,AntGrouping4}. Since each antenna port transmits the same CSI-RS, the UE is not able to distinguish the different antennas in one port and only sees a lower dimensional effective channel through CSI-RSs from the antenna ports. The UE then quantizes this lower dimensional downlink channel, which we will refer to as \textit{CSI-RS channel} throughout the paper, with a pre-defined PMI codebook and feeds back the index of selected codeword to the BS.

The 3GPP standard has defined two kinds of codebooks for efficient limited feedback, i.e., the Type 1 and Type 2 codebooks \cite{3GPP:38.214}. The Type 1 codebook is a standard PMI codebook consists of precoding matrices while the Type 2 codebook is to quantize the CSI-RS channel itself or its subspace.\footnote{Uplink feedback using the Type 2 codebook is often referred to as explicit feedback \cite{Clercks:2010}.} Although the Type 2 codebook gives better quantization performance than the Type 1 codebook, its feedback overhead increases significantly for higher layer transmission, making it unsuitable to spatial multiplexing \cite{Ahmed:2018,Ahmed:2019}. 

In addition to PMI quantization error, there are two possible factors that could result in performance degradation for spatial multiplexing: 1) usually the same beamforming weights for the CSI-RS transmissions are used for spatial multiplexing without adapting to channel conditions, and 2) the dimension of fed back PMI is usually much smaller than the dimension of original MIMO channel between the BS and the UE, which makes the BS only have very limited knowledge of the downlink MIMO channel.  These problems exist regardless of the codebook types \cite{3GPP:38.214,Ahmed:2018,Ahmed:2019}. Therefore, there is a demand on finding the full dimensional downlink MIMO channel from the low dimensional effective CSI-RS channel at the BS to maximize the performance of spatial multiplexing.

To the best of our knowledge, there has been no prior work on downlink MIMO reconstruction to support spatial multiplexing from uplink channel information.\footnote{Although \cite{Han:2019,Gao:2015,Hyuckjin:2020(FDDoverhead)} tackled similar problems, these works are limited to single antenna UEs, and it is difficult to extend the techniques in \cite{Han:2019,Gao:2015,Hyuckjin:2020(FDDoverhead)} to multiple antenna UEs as discussed before.} Since there is no relevant prior work, in this paper, we propose and compare several possible approaches for the BS to reconstruct the full downlink MIMO channel when the number of transmit antennas is less than the number of receive antennas at the UE. The proposed techniques exploit both the downlink CSI-RS and uplink SRS considering practical antenna structures to mitigate the downlink CSI-RS overhead in massive MIMO systems. The proposed techniques range from a very simple approach to complex techniques based on convex optimization problems. Among many possible approaches, we verify using the realistic spatial channel model (SCM) \cite{3GPP:36.873}, which is adopted in the 3GPP standard, that it is possible to reconstruct the downlink MIMO channel quite well using only basic matrix operations. The proposed techniques can be used for both uniform linear arrays (ULAs) and uniform planar arrays (UPAs) in a unified way. Numerical results show that the spectral efficiencies by spatial multiplexing based on the proposed downlink MIMO CSI reconstruction techniques outperform that of conventional approach, which only exploits the fed back PMI from the UE.

The remainder of this paper is organized as follows. System model and key assumptions are discussed in Section \ref{System Model}. In Section \ref{Channel Reconstruction Methods}, the proposed downlink MIMO CSI reconstruction techniques using the low dimensional effective CSI-RS channel and the uplink SRS are presented. Numerical results that verify the performance of the proposed techniques are presented in Section \ref{numerical results}, and conclusion follows in Section~\ref{conclusion}.

\textit{Notations:} Lower and upper boldface letters represent column vectors and matrices.  $\bA^{\mathrm{T}}$, $\bA^\mathrm{H}$, and $\bA^\dagger$ denote the transpose, conjugate transpose and pseudo-inverse of the matrix $\bA$. $\bA(:,m:n)$ denotes the submatrix consists of the $m$-th column to the $n$-th column of the matrix $\bA$, $\bA(m:n,:)$ denotes the submatrix consists of the $m$-th row to the $n$-th row of the matrix $\bA$, and $\ba(m:n)$ denotes the vector consists of the $m$-th element to the $n$-th element of the vector $\ba$. $\bA^\mathrm{H}(:,k)$ denotes the $k$-th column of $\bA^\mathrm{H}$, and $\bA^\mathrm{H}(k,:)$ denotes the $k$-th row of $\bA^\mathrm{H}$. $\lvert\cdot\rvert$ is used to denote the absolute value of a complex number, $\left\Vert \cdot \right\Vert$ denotes the $\ell_2$-norm of a vector, and $\left\Vert \cdot \right\Vert_{\mathrm{F}}$ denotes the Frobenius-norm of a matrix. $\mathbf{0}_m$ is used for the $m\times 1$ all zero vector, and $\bI_m$ denotes the $m \times m$ identity matrix. $\cC\cN(m,\sigma^2)$ denotes the complex normal distribution with mean $m$ and variance $\sigma^2$. $\mathcal{O}(\cdot)$ denotes Big-O notation.

\begin{figure*}[t]
	\centering
	\includegraphics[width=1.25\columnwidth]{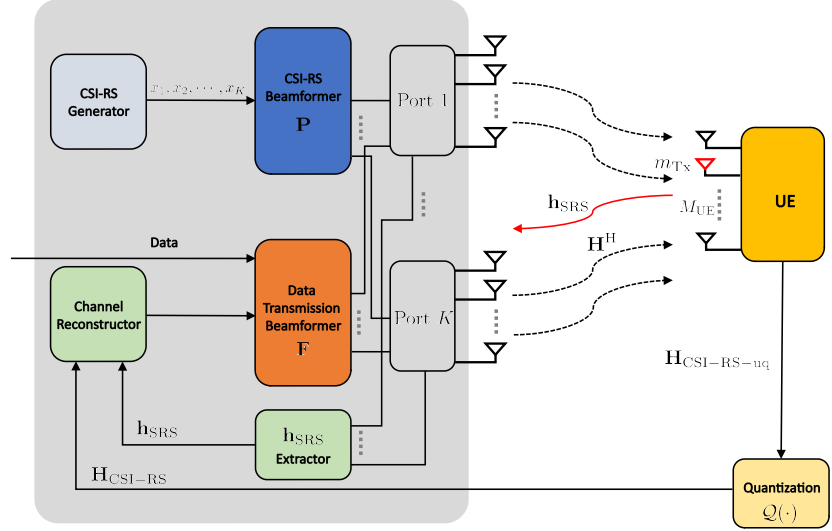} % 1.25 for two col / 0.7 for one col
	\caption{Massive MIMO with (i) downlink channel training through CSI-RS and uplink limited feedback using a PMI codebook, (ii) uplink channel training using SRS, (iii) downlink channel reconstruction. The uplink SRS channel is denoted by $\bh_{\mathrm{SRS}}$, and the index of UE antenna, which is used for transmission, is denoted by $m_{\mathrm{Tx}}$.}
	\label{system model-fig}
\end{figure*}

\section{System Model} \label{System Model} 

We consider a TDD massive MIMO system, especially SU-MIMO with spatial multiplexing. We further consider a standard PMI codebook, e.g., the Type 1 codebook, for CSI quantization.  We assume the BS is equipped with $N_\mathrm{BS}$ antennas, and the UE is equipped with $M_\mathrm{UE}$  antennas as shown in Fig. \ref{system model-fig}. At the UE side, all $M_\mathrm{UE}$ antennas are used for reception while only one of them,\footnote{It is possible to have more than one transmit antennas at the UE, and we leave this extension as a possible future work.}  indexed as $m_\mathrm{Tx}$, is used for transmission. The BS is deployed with either a ULA or a UPA while the UE is deployed with a ULA.
The overall procedure of our full downlink MIMO CSI reconstruction framework in Fig. \ref{system model-fig} is  first summarized as follows.
\begin{enumerate}[\indent Step  1:]
	\item The BS transmits beamformed CSI-RSs to the UE. Since the BS groups multiple antennas as a single port, the UE only sees low dimensional effective CSI-RS channel $\bH_{\mathrm{CSI-RS-uq}}$.
	\item The UE quantizes $\bH_{\mathrm{CSI-RS-uq}}$ using a pre-defined PMI codebook and feeds back the index of selected PMI, $\bH_{\mathrm{CSI-RS}}$, to the BS.
	\item The UE transmits uplink SRS, and the BS obtains $\bh_{\mathrm{SRS}}$ relying on the downlink and uplink channel reciprocity in TDD.
	\item Using both $\bH_{\mathrm{CSI-RS}}$ and $\bh_{\mathrm{SRS}}$, the BS reconstructs full downlink MIMO CSI.
	\item Based on the reconstructed MIMO CSI, the BS supports the UE through spatial multiplexing.
\end{enumerate}	

\begin{figure}[t]
	\centering
	\includegraphics[width=0.55\columnwidth]{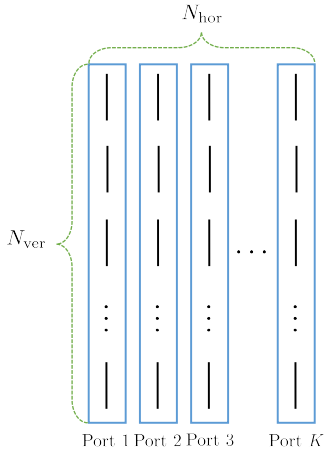} % 0.5 for two col / 0.35 for one col
	\caption{An example of practical UPA structure with $N_{\mathrm{ver}}$ vertical antennas and $N_{\mathrm{hor}}$ horizontal antennas. Each antenna port has $J=N_{\mathrm{ver}}$ antenna elements in this example.}
	\label{antenna grouping-fig}
\end{figure}

\begin{figure}[t]
	\centering
	\includegraphics[width=1\columnwidth]{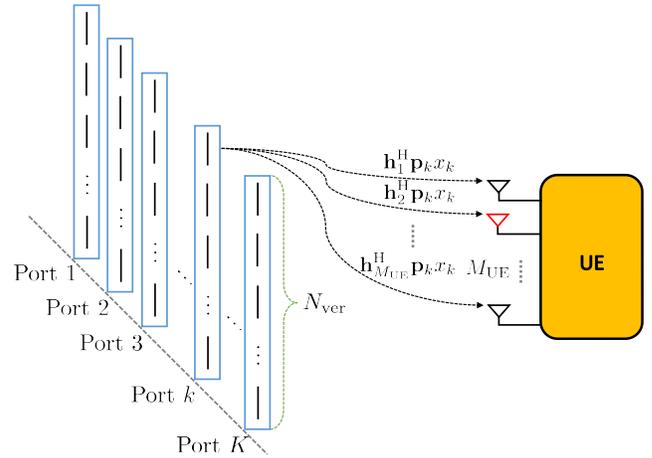} % 1 for two col / 0.6 for one col
	\caption{An example of downlink CSI-RS procedure with the antenna structure in Fig. \ref{antenna grouping-fig}. The CSI-RSs are transmitted successively from Port 1 to Port $K$ to the UE.}
	\label{signal per port-fig}
\end{figure}

As a way of mitigating the downlink CSI-RS overhead, several physical antenna elements can form a single antenna port at the BS \cite{Pi:2016magazine}. The same CSI-RS is transmitted from each antenna port so that the UE considers one antenna port as a single transmit antenna. To improve the quality of CSI-RS at the UE, however, different antenna elements in one antenna port can have different beamforming weights. When each antenna port consists of $J$ antenna elements, $K=N_\mathrm{BS}/J$ antenna ports are deployed as in Fig. \ref{antenna grouping-fig}. If the BS has prior knowledge of channel, for example through uplink SRS exploiting the channel reciprocity, the BS can dynamically select appropriate CSI-RS beamforming weights, otherwise, fixed CSI-RS beamforming weights can be applied \cite{AntGrouping1}. After constructing the CSI-RS beamforming weights, the BS transmits known CSI-RS sequences successively to the UE through the antenna ports as in Fig. \ref{signal per port-fig}. In general, the CSI-RS sequences are based on pseudo-random sequences \cite{3GPP:38.211}. Since the proposed downlink channel reconstruction techniques work for arbitrary CSI-RS sequences, as long as the BS and the UE share the same ones, we assume the CSI-RS transmitted from the $k$-th antenna port is a scalar $x_k$, not a sequence, for simplicity in this paper.

%Quantization of CSI-RS based on the received signals at the UE side is follows, then, the UE would feeds back the selected index of a PMI codebook to the BS.

Since the antenna ports transmit the CSI-RS successively, the received signals at the UE from the $k$-th antenna port $\by_k$ is given as 
\begin{equation} \label{received signal}
\by_k=\bH^\mathrm{H}\bp_k x_k + \bn_k ,
\end{equation} 
where $\bH$ is the $N_\mathrm{BS}\times M_\mathrm{UE}$ downlink MIMO channel matrix, $x_k$ is the CSI-RS  satisfying $|x_k|^2=1$, $\bn_k\sim\cC\cN(\mathbf{0}_{M_\mathrm{UE}},\sigma_\mathrm{DL}^2\bI_{M_\mathrm{UE}})$ is the $M_\mathrm{UE}\times 1$ noise vector, $\sigma_\mathrm{DL}^2=1/\rho_\mathrm{DL}$, and $\rho_\mathrm{DL}$ denotes the downlink signal-to-noise ratio (SNR). The $N_\mathrm{BS} \times 1$ CSI-RS beamforming vector $\bp_k$ is given~by
\begin{equation} \label{weight vector}
\bp_k = \left [ \mathbf{0}_{(k-1)J}^{\mathrm{T}}, \bw_k^{\mathrm{T}}, \mathbf{0}_{N_\mathrm{BS}-kJ}^{\mathrm{T}} \right]^{\mathrm{T}},
\end{equation} 
where $\bw_k$ is the $J \times 1$ CSI-RS beamforming weight vector applied to the $k$-th antenna port such that $\left\Vert \bw_k \right\Vert^2=1$. Note that all other antenna ports except the $k$-th port are silent during the $k$-th CSI-RS transmission. As discussed above, the BS can dynamically adjust $\bw_k$ if it has prior knowledge of channel, e.g., through the uplink SRS. If the BS has no prior channel knowledge, it needs to fix $\bw_k$ to have a widebeam shape to guarantee that the UE can experience a certain level of quality of service regardless of its location on its serving cell. The performance according to different CSI-RS beamforming weight vectors will be compared in Section \ref{numerical results}. 

After receiving all $K$ CSI-RS transmissions, the UE can construct an $M _\mathrm{UE}\times K$ unquantized low dimensional effective CSI-RS channel matrix $\bH_{\mathrm{CSI-RS-uq}}$. We set $x_k=1$ for simplicity throughout the paper since the proposed techniques do not rely on any CSI-RS structure.  Then $\bH_{\mathrm{CSI-RS-uq}}$ is represented by
\begin{align} 
\bH_{\mathrm{CSI-RS-uq}} 
&= \left[\by_1, \by_2, \cdots, \by_K \right],   \\
&= \bH^\mathrm{H} \bP\bX+\bN, \\
&= \bH^\mathrm{H} \bP+\bN, \label{CSI-RS-uq}
\end{align}
where $\bP=\left[\bp_1, \bp_2, \cdots, \bp_K \right], \bX= \diag \left[ x_1, x_2,\cdots, x_K \right]=\bI_K$, and $\bN=\left[\bn_1, \bn_2, \cdots, \bn_K \right]$. The UE then quantizes $\bH_{\mathrm{CSI-RS-uq}}$ using a pre-defined PMI codebook and feeds back the index of selected PMI to the BS through limited feedback. Assuming the selected PMI is for layer $L$ transmission via spatial multiplexing, the selected PMI 
\begin{equation} \label{quantization eqaution}
\bH_{\mathrm{CSI-RS}}=\cQ(\bH_{\mathrm{CSI-RS-uq}}),
\end{equation}
is a $K \times L$ matrix where $\cQ(\cdot)$ is a CSI quantization function. We consider $\cQ(\cdot)$ as a function of maximizing the spectral efficiency \cite{Love:2008}, which is given by
\begin{align} \label{quantization-max rate}
&\bH_{\mathrm{CSI-RS}} =
\notag \\  &\argmax_{\tilde{\bH}_{\mathrm{CSI-RS}} \in \cC} \log_2 \Big(\det\Big(\bI_{L}+\frac{\rho_\mathrm{DL}}{L}  \tilde{\bH}_{\mathrm{CSI-RS}}^\mathrm{H}\bH_{\mathrm{CSI-RS-uq}}^\mathrm{H}
\notag \\  & \quad\quad\quad\quad\quad\quad\quad\quad\quad\quad\quad\quad\bH_{\mathrm{CSI-RS-uq}}\tilde{\bH}_{\mathrm{CSI-RS}}\Big)\Big),
\end{align}
where $\cC$ is the pre-defined PMI codebook. Since the UE only sees the low dimensional effective channel $\bH_{\mathrm{CSI-RS-uq}}$, the UE can find the codeword that maximizes (\ref{quantization-max rate}) through exhaustive search over $\cC$. 
Note that $L$ is smaller than or equals to $M_{\mathrm{UE}}$ while the proposed downlink channel reconstruction techniques can be applied to any values of $L$.  We will show numerical results with different values of $L$ in Section \ref{numerical results}.

In TDD, relying on the downlink and uplink channel reciprocity, the BS can estimate the downlink channel from the uplink SRSs transmitted by the UE. Since we assume the UE has only one transmit antenna, the BS can estimate an $N_\mathrm{BS} \times 1$ uplink SRS channel vector $\bh_{\mathrm{SRS}}$ that corresponds to one of the columns of the downlink channel matrix $\bH$ corrupted with noise as
\begin{align} \label{SRS channel vector}
\bh_{\mathrm{SRS}}=\bH(:,m_\mathrm{Tx})+\bv,
\end{align} 
where $\bv \sim \cC\cN(\mathbf{0}_{N_\mathrm{BS}},\sigma_\mathrm{UL}^2\bI_{N_\mathrm{BS}})$ is the $N_\mathrm{BS} \times 1$ noise vector, $\sigma_\mathrm{UL}^2=1/\rho_\mathrm{UL}$, and $\rho_\mathrm{UL}$ denotes the uplink SNR.

Since the BS would not be able to know in advance which antenna is used for transmission among $M_\mathrm{UE}$ antennas at the UE, the BS does not know which column of $\bH$ corresponds to the uplink SRS channel vector.
In Section~\ref{Channel Reconstruction Methods}, we first assume $m_\mathrm{Tx}$ is known to the BS to develop downlink channel reconstruction techniques using $\bH_{\mathrm{CSI-RS}}$ and $\bh_{\mathrm{SRS}}$. Then, we show that the effect of imperfect knowledge of $m_\mathrm{Tx}$ would be negligible in terms of spectral efficiencies. We further verify the effect of $m_\mathrm{Tx}$ numerically in Section \ref{numerical results}.

\section{Proposed Downlink MIMO Channel Reconstruction Techniques} \label{Channel Reconstruction Methods}
In this section, we propose and compare possible downlink MIMO channel reconstruction techniques using the low dimensional effective CSI-RS channel and the uplink SRS. It turns out that some of proposed approaches, even complex, do not work well, i.e., even worse than the conventional spatial multiplexing only using the quantized PMI. We will still explain these approaches since there is no prior work on this problem, and readers may not know whether these approaches perform well or not.

We assume perfect knowledge of $m_\mathrm{Tx}$ at the BS. The UE feeds back the quantized $\bH_{\mathrm{CSI-RS}}$, which is selected for layer $L$ spatial multiplexing, to the BS. Since the UE has already decided that the layer $L$ spatial multiplexing would be the best for the current channel after receiving the CSI-RSs, it is reasonable to assume that the BS would reconstruct an $N_\mathrm{BS} \times L$, not $N_\mathrm{BS} \times M_\mathrm{UE}$, downlink channel if the BS obtains the layer $L$ $\bH_{\mathrm{CSI-RS}}$ from the UE. In this case, we assume $m_\mathrm{Tx}$ is in between $1$ and $L$. Note that we consider the conjugate transpose on $\bH_{\mathrm{CSI-RS}}$, i.e., $\bH^\mathrm{H}_{\mathrm{CSI-RS}}$, for the downlink channel reconstruction since the original purpose of the PMI codebook is to inform the BS of beamformer for spatial multiplexing. Then, the BS needs to take the conjugate transpose on the fed back PMI to consider it as a downlink channel.
%In general, $L$ is smaller or equal than $M_{\mathrm{UE}}$ because of practical issues. Considering the structure of PMI codebooks, however, the BS can replace the $K \times L$ selected PMI with a $K \times M_\mathrm{UE}$ codeword. For this reason, we assume $L$ and $M_{\mathrm{UE}}$ are the same through this subsection and describe the numerical results when $L$ is smaller than $M_{\mathrm{UE}}$.
\subsection{Ratio technique} \label{ratio method}
Considering $\bH=[\bh_1, \bh_2, \cdots, \bh_{M_{\mathrm{UE}}}]$, the unquantized effective CSI-RS channel matrix $\bH_{\mathrm{CSI-RS-uq}}$ in (\ref{CSI-RS-uq}) can be further represented by

% 1 col version
\begin{align} 
\bH_{\mathrm{CSI-RS-uq}}
&=
\begin{bmatrix}
\bh_1^\mathrm{H} \bp_1 & \bh_1^\mathrm{H} \bp_2 & \cdots & \bh_1^\mathrm{H} \bp_K \\
\bh_2^\mathrm{H} \bp_1 & \bh_2^\mathrm{H} \bp_2 & \cdots & \bh_2^\mathrm{H} \bp_K \\
\vdots & \vdots & \ddots & \vdots \\
\bh_{M_{\mathrm{UE}}}^\mathrm{H} \bp_1 & \bh_{M_{\mathrm{UE}}}^\mathrm{H} \bp_2 & \cdots & \bh_{M_{\mathrm{UE}}}^\mathrm{H} \bp_K \\
\end{bmatrix}
\notag\\
\notag \\
 &\quad +
\begin{bmatrix}
\bn_1, \bn_2, \cdots, \bn_K
\end{bmatrix}, \label{CSI-RS-uq-expand}
\end{align}
 
\noindent where $\bh_m$ is the $N_\mathrm{BS} \times 1$ channel vector between the transmit antennas at the BS and the $m$-th receive antenna at the UE for $m=1,2,\cdots,M_\mathrm{UE}$. Without the noise in (\ref{CSI-RS-uq-expand}), the $(m,k)$-th component of $\bH_{\mathrm{CSI-RS-uq}}$ is the inner product between $\bh_m$ and the CSI-RS beamforming vector $\bp_k$, and the only difference among components of $\bH_{\mathrm{CSI-RS-uq}}$ in the same column is $\bh_m$ with fixed $\bp_k$. Using the knowledge of $\bh_{\mathrm{SRS}}$ and $m_\mathrm{Tx}$, the BS can simply reconstruct the downlink channel in a block-wise manner with the ratio of components of $\bH_{\mathrm{CSI-RS}}$. The reconstructed $m'$-th column of downlink channel based on the \textit{ratio technique} can be expressed as
\begin{align}
\hat{\bH}_{\mathrm{ratio}}(:,&m')=\Bigg[\bh_{\mathrm{SRS}}^\mathrm{T}(1:J)\frac{\bH_{\mathrm{CSI-RS}}^\mathrm{H}({m'},1)}{\bH_{\mathrm{CSI-RS}}^\mathrm{H}(m_\mathrm{Tx},1)},\notag\\
&\bh_{\mathrm{SRS}}^\mathrm{T}(J+1:2J)\frac{\bH_{\mathrm{CSI-RS}}^\mathrm{H}({m'},2)}{\bH_{\mathrm{CSI-RS}}^\mathrm{H}(m_\mathrm{Tx},2)},\cdots,\notag\\
&\bh_{\mathrm{SRS}}^\mathrm{T}((K-1)J+1:KJ)\frac{\bH_{\mathrm{CSI-RS}}^\mathrm{H}({m'},K)}{\bH_{\mathrm{CSI-RS}}^\mathrm{H}(m_\mathrm{Tx},K)} \Bigg]^\mathrm{T},
\end{align}
for $m'=1,2,\cdots,L$. Although this technique is quite simple, it works well when the BS is deployed with the ULA as shown in Section \ref{numerical results}.

\subsection{Inner product (IP) maximization technique} \label{IP maximization method}

For the \textit{IP maximization technique}, we set an optimization problem to reconstruct the downlink channel based on the knowledge of $\bH_{\mathrm{CSI-RS}}$ and the CSI-RS beamforming
matrix $\bP$. This technique tries to maximize the IP between each row of $\bH_{\mathrm{CSI-RS}}$ and $\hat{\bH}^\mathrm{H}\bP$, which is the beamformed version of estimated downlink channel $\hat{\bH}$, as the main objective, i.e.,
\begin{equation} \label{IP-objective function}
\hat{\bH}_{\mathrm{IP}}(:,{m'})=\argmax_{\hat{\bH}(:,{m'}) \in \mathbb{C}^{N_{\mathrm{BS}} \times 1}} \left\vert(\hat{\bH}^\mathrm{H}({m'},:)\bP)\bH_{\mathrm{CSI-RS}}^\mathrm{H}(:,{m'})\right\vert, 
\end{equation}  
where $\hat{\bH}_{\mathrm{IP}}$ represents the reconstructed channel based on the \textit{IP maximization technique}.

Since the optimization in (\ref{IP-objective function}) is non-convex, we consider an equivalent convex problem as in \cite{Choi:2013}, which is given by 

\begin{align} \label{IP-objective function-2}
&\hat{\bH}_{\mathrm{IP}}(:,{m'})=
\notag\\&\argmin_{\hat{\bH}(:,{m'}) \in \mathbb{C}^{N_{\mathrm{BS}} \times 1}}\min_{\substack{\alpha\in \mathbb{R}^{+}\\ \omega\in[0,2\pi)}} \left\Vert{\hat{\bH}^\mathrm{H}({m'},:)\bP-\alpha e^{j\omega}\bH_{\mathrm{CSI-RS}}({m'},:)}\right\Vert.
\end{align}
  
\noindent After the optimization, the $m_\mathrm{Tx}$-th column of $\hat{\bH}_{\mathrm{IP}}$ is replaced by $\bh_{\mathrm{SRS}}$, i.e.,
\begin{equation} \label{IP-replacement}
\hat{\bH}_{\mathrm{IP}}(:,m_\mathrm{Tx})=\bh_{\mathrm{SRS}},
\end{equation}
since $\bh_{\mathrm{SRS}}$ is, with sufficiently high uplink SNR, close to the true channel relying on the downlink and uplink channel reciprocity while the $m_\mathrm{Tx}$-th column of $\hat{\bH}_{\mathrm{IP}}$ is the estimated~one.
 
\subsection{Element-wise technique} \label{element-wise method}
In the \textit{element-wise technique}, we set a convex optimization problem to minimize the error between $\hat{\bH}^\mathrm{H}\bP$ and $\bH_{\mathrm{CSI-RS}}$~as

\begin{align} \label{element-objective fucntion}
\hat{\bH}_{\mathrm{ele}}=\argmin_{\hat{\bH}\in \mathbb{C}^{N_\mathrm{BS} \times L }}&{\left\Vert \hat{\bH}^\mathrm{H}\bP-\bH_{\mathrm{CSI-RS}}^\mathrm{H}\right\Vert_{\mathrm{F}}}
\notag\\&+\lambda{\left\Vert\hat{\bH}(:,m_\mathrm{Tx})-\bh_{\mathrm{SRS}}\right\Vert},
\end{align}
where $\lambda \in \mathbb{R}^\mathrm{+}$ denotes the regularization factor, and $\hat{\bH}_{\mathrm{ele}}$ represents the reconstructed downlink channel based on the \textit{element-wise technique}. Large $\lambda$ implies large emphasis on minimizing the difference between the $m_\mathrm{Tx}$-th column of the reconstructed downlink MIMO channel and the known $\bh_{\mathrm{SRS}}$. It is not always better to have large $\lambda$, instead, it should be properly adjusted to balance the two differences. Similar to (\ref{IP-replacement}), the $m_\mathrm{Tx}$-th column of $\hat{\bH}_{\mathrm{ele}}$ is replaced by $\bh_{\mathrm{SRS}}$ after the optimization.

Although the \textit{IP maximization technique} and \textit{element-wise technique} exploit given information of $\bH_{\mathrm{CSI-RS}}$, $\bh_{\mathrm{SRS}}$, and $\bP$, they do not exploit any physical structure, e.g., angle-of-arrival (AoA) and angle-of-departure (AoD) of the channel $\bH$ or antenna array at the BS and UE, making them have poor performance as shown in Section \ref{numerical results}. In addition, the dimension of ${\bH}$ is quite large in massive MIMO with large $N_\mathrm{BS}$, resulting in high degree-of-freedom with only a few known variables for the optimization problems. In what follows, we impose physical structures on the optimization problems to improve reconstruction performance and mitigate optimization complexity.

\subsection{Structure technique} \label{structure method}

In massive MIMO with a large number of antennas, the downlink channel $\bH$ is usually modeled as virtual channel representation \cite{Mo:2018,Sayeed:2002}, which is the weighted sum of the outer products of AoA array response vectors at the UE side and the AoD array response vectors at the BS side. The modeled channel is given by
\begin{equation} \label{structure-model}
\bH_\mathrm{v}^{\mathrm{H}}=\sum_{p=1}^{P}\sum_{q=1}^{Q}c_{p,q} \ba_{\mathrm{r}}(\psi_p) \ba_{\mathrm{t}}^{\mathrm{H}}(\mu_q),
\end{equation}
where $c_{p,q}\in \mathbb{C}$ is the complex gain of the path with the AoD $\mu_q$ and AoA $\psi_p$. Since the UE deploys the ULA, the array response vector for AoA $\psi_p$, assuming half wavelength antenna spacing, is represented by
\begin{align}
\ba_\mathrm{r}(\psi_p) =\frac{1}{\sqrt{L}} \left[1, e^{j\pi \sin(\psi_p)}, \cdots, e^{j(L-1)\pi\sin(\psi_p)} \right]^\mathrm{T}. \label{array response ULA receiver}
\end{align}
The UE deploys $M_\mathrm{UE}$ physical receive antennas; however, we model the antenna array of the UE as deploying $L$ antennas to reconstruct the $N_\mathrm{BS} \times L$ downlink channel at the BS. Similarly, the array response vector for AoD $\mu_q$, assuming the ULA at the BS with half wavelength antenna spacing, is given as
\begin{align}
\ba_\mathrm{t}(\mu_q) =\frac{1}{\sqrt{N_\mathrm{BS}}} \left[1, e^{j\pi \sin(\mu_q)}, \cdots, e^{j(N_\mathrm{BS}-1)\pi\sin(\mu_q)} \right]^\mathrm{T}. \label{array response ULA transmitter}
\end{align}
Although the UPA is also possible, we only consider the ULA at the BS for the \textit{structure technique}, the reason will become clear at the end of this subsection.

In a matrix form, the modeled channel in (\ref{structure-model}) can be rewritten as
\begin{align} \label{structure-matrix model}
\bH_\mathrm{str}^\mathrm{H}=\bA_{\mathrm{r}} \bC \bA_{\mathrm{t}}^\mathrm{H},
\end{align}
where $\bC$ is the $P\times Q$ matrix with $c_{p,q}$ as the $(p,q)$-th element, $\bA_{\mathrm{r}}=\left[ \ba_{\mathrm{r}}(\psi_1), \ba_{\mathrm{r}}(\psi_2), \cdots, \ba_{\mathrm{r}}(\psi_P)\right]$, and $\bA_{\mathrm{t}}=\left[ \ba_{\mathrm{t}}(\mu_1), \ba_{\mathrm{t}}(\mu_2), \cdots, \ba_{\mathrm{t}}(\mu_Q)\right]$. Without any prior knowledge of AoAs and AoDs of the channel, $\psi_p$ and $\mu_q $ can be chosen randomly from $[-\pi/2,\pi/2]$ considering practical cell structures.

To reconstruct the downlink channel assuming the channel structure expressed in (\ref{structure-matrix model}) and randomly chosen AoAs and AoDs, the convex optimization problem in (\ref{element-objective fucntion}) now becomes
% 1 col
\begin{align} \label{structure-objective function}
\hat{\bC}_{\mathrm{str}}=\argmin_{\hat{\bC}\in \mathbb{C}^{P \times Q }}&{\left\Vert \bA_{\mathrm{r}} \hat{\bC} \bA_{\mathrm{t}}^\mathrm{H}\bP-\bH_{\mathrm{CSI-RS}}^\mathrm{H}\right\Vert_{\mathrm{F}}}
\notag \\&+\lambda{\left\Vert \left[\bA_{\mathrm{t}} \hat{\bC}^\mathrm{H} \bA_{\mathrm{r}}^\mathrm{H}\right](:,m_\mathrm{Tx})-\bh_{\mathrm{SRS}}\right\Vert}, 
\end{align}
\noindent where $\lambda \in \mathbb{R}^\mathrm{+}$ denotes the regularization factor as in (\ref{element-objective fucntion}). The reconstructed channel based on the \textit{structure technique} then is given as
\begin{equation}
\hat{\bH}^\mathrm{H}_{\mathrm{str}}= \bA_{\mathrm{r}} \hat{\bC}_{\mathrm{str}} \bA_{\mathrm{t}}^\mathrm{H}.
\end{equation}
Similar to (\ref{IP-replacement}), the $m_\mathrm{Tx}$-th column of $\hat{\bH}_{\mathrm{str}}$ is replaced by $\bh_{\mathrm{SRS}}$ after the optimization. 

The \textit{structure technique} may suffer from randomly chosen $\psi$'s and $\mu$'s, which could be misaligned with the true AoAs and AoDs. It is possible to resolve this problem by increasing the size of $P$ and $Q$ but this would impose huge complexity on the optimization in (\ref{structure-objective function}). With the UPA at the BS, the complexity issue becomes even worse since the BS needs to take both the horizontal and vertical angles into account. Therefore, we only consider the ULA at the BS for the \textit{structure technique}.

\subsection{Pre-search technique} \label{pre-search method}
As a way of resolving the complexity issue in the \textit{structure technique}, we first estimate the dominant AoDs and AoAs as preliminary information relying on the channel model (\ref{structure-model}) in the \textit{pre-search technique}. Since the BS has $\bH_{\mathrm{CSI-RS}}$ by limited feedback from the UE, the dominant AoAs can be extracted by comparing the strengths $\chi^{\mathrm{r}}(\tilde{\psi}_{i})$ of the arrival angle $\tilde{\psi}_{i}$ as 
\begin{align}
\chi^{\mathrm{r}}(\tilde{\psi}_{i}) &= \left\Vert \ba_{\mathrm{AoA}}^\mathrm{H}(\tilde{\psi}_{i})\bH_{\mathrm{CSI-RS}}^\mathrm{H}  \right\Vert, \\
\tilde{\psi}_{i} &= -\frac{\pi}{2}+\frac{\pi}{R_{\mathrm{ULA}}}(i-1),
\end{align}
where $i=1,2,\cdots,R_\mathrm{ULA}+1$. Here, $\pi/R_{\mathrm{ULA}}$ represents the resolution of $\tilde{\psi}_{i}$, and $\ba_{\mathrm{AoA}}(\cdot)$ is the same as $\ba_{\mathrm{r}}(\cdot)$ in (\ref{array response ULA receiver}) since the UE is assumed to deploy the ULA. To extract $T_\mathrm{AoA}$ dominant angles for AoAs, we need to find $T_\mathrm{AoA}$ local maxima of $\chi^{\mathrm{r}}(\tilde{\psi}_{i})$ where a local maximum is defined as
\begin{equation}
\chi^{\mathrm{r}}(\tilde{\psi}_{i}) \geq \chi^{\mathrm{r}}(\tilde{\psi}_{i+1}), \enspace  \chi^{\mathrm{r}}(\tilde{\psi}_{i}) \geq \chi^{\mathrm{r}}(\tilde{\psi}_{i-1}).
\end{equation}
We denote an angle that gives a local maximum of $\chi^{\mathrm{r}}(\tilde{\psi}_{i})$ as $\hat{\psi}_u$ for $u=1,2,\cdots,T_{\mathrm{AoA}}$. Note that the PMI codebook is pre-defined; therefore, it is possible to construct a lookup table that defines $T_{\mathrm{AoA}}$ dominant AoAs for each $\bH_{\mathrm{CSI-RS}}$ in~advance.

Estimating $T_{\mathrm{AoD}}$ dominant AoDs can be conducted similarly using $\bh_{\mathrm{SRS}}$. Different from the \textit{structure technique}, now it is possible to consider both the ULA and UPA for the BS antenna structure. Assuming the ULA at the BS, the strength $\chi^{\mathrm{t}}(\tilde{\mu}_i)$ of the departure angle $\tilde{\mu}_i$ is given by
\begin{align} \label{chi-AoD,ULA}
\chi^{\mathrm{t}}(\tilde{\mu}_i) &= \left\vert \ba_{\mathrm{AoD}}^\mathrm{H}(\tilde{\mu}_i) \bh_{\mathrm{SRS}} \right\vert, \\
\tilde{\mu}_i &= -\frac{\pi}{2}+\frac{\pi}{R_{\mathrm{ULA}}}(i-1),
\end{align}
where $\ba_{\mathrm{AoD}}(\cdot)$ is the same as $\ba_{\mathrm{t}}(\cdot)$ in (\ref{array response ULA transmitter}). If the UPA is assumed at the BS, the strength $\chi^{\mathrm{t}}({\tilde{\mu}_{\ell_{\mathrm{ver}}},\tilde{\mu}_{\ell_{\mathrm{hor}}}})$  of the vertical departure angle $\tilde{\mu}_{\ell_{\mathrm{ver}}}$ and horizontal departure angle $\tilde{\mu}_{\ell_{\mathrm{hor}}}$ are written as
\begin{align} \label{chi-AoD,UPA}
\chi^{\mathrm{t}}({\tilde{\mu}_{\ell_{\mathrm{ver}}},\tilde{\mu}_{\ell_{\mathrm{hor}}}}) &= \left\vert \ba_{\mathrm{AoD}}^\mathrm{H}(\tilde{\mu}_{\ell_{\mathrm{ver}}},\tilde{\mu}_{\ell_{\mathrm{hor}}}) \bh_{\mathrm{SRS}} \right\vert,
\end{align}
where the array response vector for the UPA, assuming half wavelength spacing, is given as 
\begin{align} \label{UPA-array response vector}
&\ba_{\mathrm{AoD}}(\tilde{\mu}_{\ell_{\mathrm{ver}}},\tilde{\mu}_{\ell_{\mathrm{hor}}})=
\notag \\
&\quad \quad \frac{1}{\sqrt{N_\mathrm{BS}}} \left[1, e^{j\pi \sin(\tilde{\mu}_{\ell_{\mathrm{ver}}})}, \cdots, e^{j(N_{\mathrm{ver}}-1)\pi \sin(\tilde{\mu}_{\ell_{\mathrm{ver}}})}\right]^\mathrm{T}
\notag \\
&\quad \quad \quad \enspace \enspace \otimes \Big[1, e^{j\pi \sin(\tilde{\mu}_{\ell_{\mathrm{hor}}})\cos(\tilde{\mu}_{\ell_{\mathrm{ver}}})}, \cdots, \notag \\
&\quad \quad \quad \quad \quad \quad \quad \quad \quad e^{j(N_{\mathrm{hor}}-1)\pi \sin(\tilde{\mu}_{\ell_{\mathrm{hor}}})\cos(\tilde{\mu}_{\ell_{\mathrm{ver}}})}\Big]^\mathrm{T}.
\end{align}

\noindent In (\ref{UPA-array response vector}), $N_\mathrm{BS}=N_{\mathrm{ver}} N_{\mathrm{hor}}$, and $\otimes$ denotes the Kronecker product. Further, $\tilde{\mu}_{\ell_{\mathrm{ver}}}$ and $\tilde{\mu}_{\ell_{\mathrm{hor}}}$ are conditioned by
\begin{align}
\tilde{\mu}_{\ell_{\mathrm{ver}}} &= -\frac{\pi}{2}+\frac{\pi}{R_{\mathrm{UPA,ver}}}(\ell_{\mathrm{ver}}-1), \\
\ell_{\mathrm{ver}} &= 1,2,\cdots, R_{\mathrm{UPA,ver}}+1, \\
\tilde{\mu}_{\ell_{\mathrm{hor}}} &= -\frac{\pi}{2}+\frac{\pi}{R_{\mathrm{UPA,hor}}}(\ell_{\mathrm{hor}}-1), \\
\ell_{\mathrm{hor}} &= 1,2,\cdots, R_{\mathrm{UPA,hor}}+1,
\end{align}
where $\pi/R_{\mathrm{UPA,ver}}$ and $\pi/R_{\mathrm{UPA,hor}}$ represent the resolution of $\tilde{\mu}_{\ell_{\mathrm{ver}}}$ and $\tilde{\mu}_{\ell_{\mathrm{hor}}}$.

Rather than just choosing the dominant $T_\mathrm{AoD}$ departure angles, it is possible to increase the AoD estimation accuracy by the null space projection technique as in \cite{Choi:2019}. Once a dominant AoD is found, we can exclude the component corresponding to that angle from $\bh_{\mathrm{SRS}}$ through the null space projection technique before searching another dominant AoD. The details are summarized in Algorithms 1 and 2 for the ULA and UPA cases. We denote the sets of dominant AoDs for the ULA and UPA cases as $\bO_{\mathrm{ULA}}$ and $\bO_{\mathrm{UPA}}$ in those two algorithms.

After obtaining the dominant AoAs and AoDs, we have the $L \times T_{\mathrm{AoA}}$ matrix $\bA_{\mathrm{AoA}}$ given as
\begin{equation}
\bA_{\mathrm{AoA}}=\left[\ba_{\mathrm{AoA}}(\hat{\psi}_{1}),\ba_{\mathrm{AoA}}(\hat{\psi}_{2}), \cdots, \ba_{\mathrm{AoA}}(\hat{\psi}_{T_{\mathrm{AoA}}})\right],
\end{equation}
and the $N_\mathrm{BS} \times T_{\mathrm{AoD}}$ matrix $\bA_{\mathrm{AoD}}$ expressed as
\begin{equation}
\bA_{\mathrm{AoD}}=\Big[\ba_{\mathrm{AoD}}(\hat{\mu}_1),\ba_{\mathrm{AoD}}(\hat{\mu}_2),\cdots,\ba_{\mathrm{AoD}}(\hat{\mu}_{T_{\mathrm{AoD}}})\Big],
\end{equation}
assuming the ULA at the BS or
\begin{align}
\bA_{\mathrm{AoD}}=\Big[&\ba_{\mathrm{AoD}}(\hat{\mu}_{1,\mathrm{ver}},\hat{\mu}_{1,\mathrm{hor}}),\ba_{\mathrm{AoD}}(\hat{\mu}_{2,\mathrm{ver}},\hat{\mu}_{2,\mathrm{hor}}),\notag\\
&\cdots,\ba_{\mathrm{AoD}}(\hat{\mu}_{T_{\mathrm{AoD}},\mathrm{ver}},\hat{\mu}_{T_{\mathrm{AoD}},\mathrm{hor}})\Big],
\end{align}
assuming the UPA at the BS. To reconstruct the downlink channel, we set a convex optimization problem to find the path gain matrix $\hat{\bC}_{\mathrm{pre}}$ with the given $\bA_{\mathrm{AoA}}$ and $\bA_{\mathrm{AoD}}$ as 

\begin{align} \label{pre-search-objective function}
\hat{\bC}_{\mathrm{pre}}=\argmin_{\tilde{\bC}\in \mathbb{C}^{T_{\mathrm{AoA}} \times T_{\mathrm{AoD}} }}{\left\Vert \bA_{\mathrm{AoA}} \tilde{\bC} \bA_{\mathrm{AoD}}^\mathrm{H}\bP-\bH_{\mathrm{CSI-RS}}^\mathrm{H}\right\Vert_{\mathrm{F}}} \notag \\
+\lambda{\left\Vert \left[ \bA_{\mathrm{AoD}} \tilde{\bC}^\mathrm{H} \bA_{\mathrm{AoA}}^\mathrm{H}\right](:,m_\mathrm{Tx})-\bh_{\mathrm{SRS}}\right\Vert},
\end{align}
\noindent where $\lambda \in \mathbb{R}^\mathrm{+}$ denotes the regularization factor as in (\ref{element-objective fucntion}). The reconstructed channel $\hat{\bH}_{\mathrm{pre}}$ based on the \textit{pre-search technique} is given as
\begin{align} \label{pre-search-reconstructed}
\hat{\bH}_{\mathrm{pre}}^\mathrm{H}=\bA_{\mathrm{AoA}} \hat{\bC}_{\mathrm{pre}} \bA_{\mathrm{AoD}}^\mathrm{H}.
\end{align}
Similar to (\ref{IP-replacement}), the $m_\mathrm{Tx}$-th column of $\hat{\bH}_{\mathrm{pre}}$ is replaced by $\bh_{\mathrm{SRS}}$ after the optimization. Note that the \textit{pre-search technique} has lower complexity for optimization than the \textit{structure technique} since $T_{\mathrm{AoA}}$ and $T_{\mathrm{AoD}}$ would be smaller than $P$ and $Q$ to have the same performance in general. Even though the size of optimization problem has become smaller, still it might take much time to perform the \textit{pre-search technique} in practice, which could prevent its use when the channel coherence time is insufficient.

\alglanguage{pseudocode}
\begin{algorithm}[t]
	\caption{Estimation of the dominant AoDs for the ULA}
	\begin{algorithmic}
		\State Initialize $\bO_{\mathrm{ULA}}$ as an empty set
		\State $\bh \leftarrow \bh_{\mathrm{SRS}}$
		\For {$v=1,2,\cdots,T_{\mathrm{AoD}}$}~
		\State Initialize $i_\mathrm{max}$
		\For{$i=1,2,\cdots,R_{\mathrm{ULA}}+1$}
		\State Calculate $\chi^{\mathrm{t}}(\tilde{\mu}_i)$ in (\ref{chi-AoD,ULA}) 
		%		\State Calculate $\chi^{\mathrm{t}}_i$ 
		\EndFor
		\State Calculate $i_{\mathrm{max}}={\underset{i}{\argmax}} \enspace \chi^{\mathrm{t}}(\tilde{\mu}_i)$
		%		\State Update 
		\State $\hat{\mu}_v \leftarrow \tilde{\mu}_{i_\mathrm{max}}$
		\State  $\bh \leftarrow \bh - (\bh^\mathrm{H}\ba_{\mathrm{AoD}}(\hat{\mu}_v))\ba_{\mathrm{AoD}}(\hat{\mu}_v)$
		\State $\bO_{\mathrm{ULA}} \leftarrow \{\bO_{\mathrm{ULA}},\hat{\mu}_v\}$
		\EndFor
	\end{algorithmic}
\end{algorithm}

\subsection{Pseudo-inverse technique} \label{pseudo-inverse method}

All the above techniques except the \textit{ratio technique} consider certain convex optimization problems for which the convergence is guaranteed. However, the overall complexity to reconstruct the downlink channel through an optimization problem can be quite severe especially in massive MIMO. We propose another channel reconstruction technique that only exploits basic matrix operations and does not rely on any optimization to combat the complexity problem. This would be especially beneficial when the channel coherence time is not long enough to perform any complex optimization process.

Adopting the channel model as in (\ref{pre-search-reconstructed}), $\bH_{\mathrm{CSI-RS-uq}}$ in (\ref{CSI-RS-uq}) can be represented by
\begin{align}
\bH_{\mathrm{CSI-RS-uq}}
&=\bH^\mathrm{H}\bP+\bN,  \\ 
&\approx \bA_{\mathrm{AoA}} \tilde{\bC} \bA_{\mathrm{AoD}}^\mathrm{H} \bP+\bN,
\end{align}
where $\bA_{\mathrm{AoA}}$ and $\bA_{\mathrm{AoD}}$ are obtained by the same way as in the \textit{pre-search technique}. Note that $T_{\mathrm{AoA}}$ and $T_{\mathrm{AoD}}$ for finding the dominant AoAs and AoDs are design variables that the BS can choose. By setting $T_{\mathrm{AoA}} \leq L$ and $T_{\mathrm{AoD}} \leq K$, the left pseudo-inverse of $\bA_{\mathrm{AoA}}$ and the right pseudo-inverse of $\bA_{\mathrm{AoD}}^\mathrm{H} \bP$ always exist. Then, the estimated  $\hat{\bC}_{\mathrm{pinv}}$ is given by
\begin{equation} \label{pseudo-inverse channel gain}
\hat{\bC}_{\mathrm{pinv}} = \bA^{\dagger}_{\mathrm{AoA}} \bH_{\mathrm{CSI-RS}}^\mathrm{H} (\bA_{\mathrm{AoD}}^\mathrm{H} \bP)^\dagger,
\end{equation}
and the reconstructed channel $\hat{\bH}_{\mathrm{pinv}}$ based on the \textit{pseudo-inverse technique} is given as
\begin{equation}
\hat{\bH}_{\mathrm{pinv}}=\bA_{\mathrm{AoA}} \hat{\bC}_{\mathrm{pinv}} \bA_{\mathrm{AoD}}^\mathrm{H}.
\end{equation}
Similar to (\ref{IP-replacement}), the $m_\mathrm{Tx}$-th column of $\hat{\bH}_{\mathrm{pinv}}$ is replaced by~$\bh_{\mathrm{SRS}}$ after the reconstruction.

\alglanguage{pseudocode}
\begin{algorithm}[t]
	\caption{Estimation of the dominant AoDs for the UPA}
	\begin{algorithmic}
		\State Initialize $\bO_{\mathrm{UPA}}$ as an empty set
		\State $\bh \leftarrow \bh_{\mathrm{SRS}}$
		\For {$v=1,2,\cdots,T_{\mathrm{AoD}}$}~
		\State Initialize $\ell_{\mathrm{ver,max}},\ell_{\mathrm{hor,max}}$
		\For{$\ell_{\mathrm{ver}}=1,2,\cdots,R_{\mathrm{UPA,ver}}+1$}
		\For{$\ell_{\mathrm{hor}}=1,2,\cdots,R_{\mathrm{UPA,hor}}+1$}
		\State Calculate $\chi^{\mathrm{t}}({\tilde{\mu}_{\ell_{\mathrm{ver}}},\tilde{\mu}_{\ell_{\mathrm{hor}}}})$ in (\ref{chi-AoD,UPA}) 
		%		\State Calculate $\chi^{\mathrm{t}}_{\ell_{\mathrm{ver}},\ell_{\mathrm{hor}}}$ 
		\EndFor
		\EndFor
		\State Calculate $\left( \ell_{\mathrm{ver,max}},\ell_{\mathrm{hor,max}}\right)={\underset{\ell_{\mathrm{ver}},\ell_{\mathrm{hor}}}{\argmax}} \chi^{\mathrm{t}}({\tilde{\mu}_{\ell_{\mathrm{ver}}},\tilde{\mu}_{\ell_{\mathrm{hor}}}})$
		%		\State Update 
		\State $\hat{\mu}_{v,\mathrm{ver}} \leftarrow \tilde{\mu}_{\ell_{\mathrm{ver,max}}}$ 
		\State $\hat{\mu}_{v,\mathrm{hor}} \leftarrow \tilde{\mu}_{\ell_{\mathrm{hor,max}}}$
		\State $\bh \leftarrow \bh - (\bh^\mathrm{H}\ba_{\mathrm{AoD}}(\hat{\mu}_{v,\mathrm{ver}},\hat{\mu}_{v,\mathrm{hor}}))\ba_{\mathrm{AoD}}(\hat{\mu}_{v,\mathrm{ver}},\hat{\mu}_{v,\mathrm{hor}})$
		\State $\bO_{\mathrm{UPA}} \leftarrow \{\bO_{\mathrm{UPA}},(\hat{\mu}_{v,\mathrm{ver}},\hat{\mu}_{v,\mathrm{hor}})\}$
		\EndFor
	\end{algorithmic}
\end{algorithm}

\subsection{Complexity analysis}
Among the proposed techniques, the \textit{IP maximization, element-wise, structure} and \textit{pre-search techniques} need to solve the convex optimization problems. Although these problems can be efficiently solved using the interior-point method, its complexity is incomparable to the complexity of basic matrix-vector operations. On the contrary, the \textit{ratio} and \textit{pseudo-inverse techniques} solely rely on the basic matrix-vector operations. Specifically, the complexity of \textit{ratio technique} is $\mathcal{O}(N_{\mathrm{BS}}L)$ since it needs to obtain the inner product of two vectors $L$ times. The \textit{pseudo-inverse technique} requires to have the dominant AoA/AoD information where the complexity of AoA estimation based on $\bH_{\mathrm{CSI-RS}}$ is $\mathcal{O}(LKR_{\mathrm{ULA}})$, and that of AoD estimation using  $\bh_{\mathrm{SRS}}$ is $\mathcal{O}(N_{\mathrm{BS}}R_{\mathrm{ULA}})$ for the ULA and  $\mathcal{O}(N_{\mathrm{BS}}R_{\mathrm{UPA,hor}}R_{\mathrm{UPA,ver}})$ for the UPA at the BS. The complexity of channel gain matrix estimation in (\ref{pseudo-inverse channel gain}) is $\mathcal{O}(L T_{\mathrm{AoA}}^2 + K^3 + T_{\mathrm{AoD}}N_{\mathrm{BS}}K +LK T_{\mathrm{AoA}})$. Although the complexity of \textit{pseudo-inverse technique} is higher than that of \textit{ratio technique}, it is only proportional to $N_\mathrm{BS}$ and much lower than the complexity of interior-point method.

\subsection{Effect of imperfect knowledge of transmit antenna index of UE at BS} \label{effect of m_Tx}

Until now, we assumed the BS has the perfect knowledge of $m_\mathrm{Tx}$, i.e., the transmit antenna index of the UE, to reconstruct the downlink channel. The BS, however, may have imperfect knowledge about $m_\mathrm{Tx}$ in practice. To see the effect of imperfect knowledge of $m_\mathrm{Tx}$ on the spectral efficiency performance, we first assume $L$ is the same as $M_\mathrm{UE}$ for simplicity. Then, the spectral efficiency of channel is defined as \cite{Love:2008} 
\begin{align}
R&=\log_2\left(\det\left(\bI_{M_\mathrm{UE}}+\frac{\rho_\mathrm{DL}}{M_\mathrm{UE}}\bF^\mathrm{H}\bH\bH^\mathrm{H}\bF\right)\right), \label{rate} \\ 
\bF&={\bV}(:,1:M_\mathrm{UE}), \label{beamforemr} \\
{\bH}^\mathrm{H}&= {\bU} {\Sigma} {\bV}^\mathrm{H}, \label{SVD} 
\end{align}
where (\ref{SVD}) is the singular value decomposition (SVD) of the true downlink channel ${\bH}^\mathrm{H}$, and $\bF$ is the optimal data transmission beamformer. Let $\bT$ be an arbitrary row permutation matrix. Then the SVD on the row permuted downlink channel $\bT{\bH}^\mathrm{H}$ is given as
\begin{equation}
\bT{\bH}^\mathrm{H}= \left(\bT{\bU}\right) {\Sigma} {\bV}^\mathrm{H},
\end{equation}
where $\bT{\bU}$ is still a unitary matrix. Since the right singular matrix $\bV$ is not altered by $\bT$, the spectral efficiency becomes the same regardless of $\bT$. 

In our downlink channel reconstruction problem, the imperfect knowledge of $m_\mathrm{Tx}$ works as the row permutation matrix~$\bT$. Of course incorrect knowledge of $m_\mathrm{Tx}$ would result in a different reconstruction result in addition to the row permutation effect. 
It is difficult, however, to analytically derive the impact of imperfect knowledge of $m_\mathrm{Tx}$ on the downlink MIMO CSI reconstruction. Therefore, we numerically study this impact in Section \ref{numerical results} where the result shows that the imperfect knowledge of $m_\mathrm{Tx}$ has negligible impact on the spectral efficiency performance. This information could be important for practical implementation, e.g., symbol detection at the UE, which could be an interesting future research topic.

\section{Numerical Results} \label{numerical results}

In this section, we evaluate the performance of the proposed downlink channel reconstruction techniques. The downlink channel $\bH$ is generated based on the SCM channel that is extensively used in the 3GPP standard \cite{3GPP:36.873}. Unless explicitly stated, we adopt the scenario of urban micro (UMi) single cell with carrier frequency 2.3 GHz for the SCM channel. Since the SCM channel takes cell structures with path loss into account, channel gains are usually very small. For numerical studies of point-to-point communication using spatial multiplexing, we normalize the average gain of all channel elements to one, i.e., $\mathbb{E}\left[\left|h_{n,m}\right|^2\right]=1$ where $h_{n,m}$ is the $(n,m)$-th component of $\bH$. 
%\begin{table*}
%	\renewcommand{\arraystretch}{2} % 2column=2
%	\centering
%	\caption{Average spectral efficiency [bps/Hz] according to the reconstruction techniques with the fixed widebeam weight at the BS and $L=M_\mathrm{UE}=4$ layer feedback. }
%	\begin{tabular}{c| c c c c c c}
%		\multirow{2}*{\shortstack{BS \\ Antenna \\ Structure}}
%		& CSI-RS & Random & Ratio & IP-max & Element  & Structure\\
%		&\centering{Pre-ULA}  & Pre-UPA  & Pinv-ULA  & Pinv-UPA  & Ideal & \\
%		\hline\hline
%		\multirow{2}*{ULA} 
%		& 22.5306  & 20.6037 & 25.5743 & 20.9157 & 20.9157  & 17.4140 \\
%		& \textbf{27.2573} & 26.9630 & 27.2349 & 26.9711 & \textbf{31.3884}  &\\
%		\hline
%		\multirow{2}{*}{UPA}
%		& 16.5176 & 22.3962 & 20.8487 & 21.6025 & 21.6025  & 22.2037 \\
%		& 26.6532 & 27.6892 & 26.7251 & \textbf{27.7944} & \textbf{34.2036} &\\
%		\hline
%	\end{tabular}
%	\label{ULA,UPA,widebeam average rate}
%\end{table*}

We set the number of transmit antennas at the BS $N_\mathrm{BS}=32$ (for the UPA, $N_{\mathrm{ver}}=8, N_{\mathrm{hor}}=4$), the number of receive antennas at the UE $M_\mathrm{UE}=4$, the number of antenna elements for an antenna port $J=8$, which gives the number of antenna ports $K=4$, and the number of PMI feedback layer $L=2$ or $L=M_\mathrm{UE}=4$. The regularization factor is numerically optimized and set as $\lambda=0.5$ for all optimization problems. We use CVX \cite{CVX}, a well established optimization solver, for some of proposed approaches that need to solve convex optimization problems. We also set the number of randomly selected angles $P=Q=20$ for the \textit{structure technique}, the number of dominant AoAs or AoDs $T_\mathrm{AoA}=L, T_\mathrm{AoD}=K$ and the resolution for finding dominant AoAs or AoDs $R_\mathrm{ULA}=3600, R_{\mathrm{UPA,ver}}= R_{\mathrm{UPA,hor}}=200$ for the \textit{pre-search technique} and \textit{pseudo-inverse technique}. The downlink SNR $\rho_\mathrm{DL}$ is assumed to be 20 dB since the spatial multiplexing is intended to increase the spectral efficiency in high SNR~regimes.

For the CSI-RS beamforming weight vector $\bw_k$ in (\ref{weight vector}), we consider a widebeam or dynamically selected beam based on $\bh_{\mathrm{SRS}}$. Specially, for the case of dynamically selected beam, we assume $\bw_{j_\mathrm{max}}$ is used where $j_{\mathrm{max}}$ is the column index of $J \times J$ discrete Fourier transform (DFT) matrix $\bD$ selected as
\begin{equation}
j_\mathrm{max}= \argmax_{j} \left| \left[\bD^\mathrm{H}(j,:),\mathbf{0}_{N_{\mathrm{BS}}-J}^\mathrm{T}\right] \bh_{\mathrm{SRS}} \right|. \label{j_max}
\end{equation} 
Then, $\bw_{j_\mathrm{max}}$ is defined by the $j_{\mathrm{max}}$-th column of $\bD$. Note that $\bw_k$ may vary depending on $k$ in general; however, we assume those are the same for all $k$. For the PMI codebook~$\cC$, the Type 1 Single-Panel Codebook in \cite{3GPP:38.214} is adopted. As a performance metric, we consider the spectral efficiency of the channel with reconstructed downlink channel replacing $M_\mathrm{UE}$ with $L$ in (\ref{rate}). The data transmission beamformer $\bF$ is set as  
\begin{align}
\bF&=\hat{\bV}(:,1:L), \\
\hat{\bH}^\mathrm{H}&= \hat{\bU} \hat{\Sigma} \hat{\bV}^\mathrm{H}, \label{SVD-hat}
\end{align}
where (\ref{SVD-hat}) is the SVD of the reconstructed channel $\hat{\bH}^\mathrm{H}$ by the downlink channel reconstruction techniques proposed in Section \ref{Channel Reconstruction Methods}.

In the following figures, the term Pre-ULA (Pre-UPA) refers to the \textit{pre-search technique} explained in Section \ref{pre-search method} with the ULA (UPA) assumption at the BS, and Pinv-ULA (Pinv-UPA) refers to the \textit{pseudo-inverse technique} in Section \ref{pseudo-inverse method} with the ULA (UPA) assumption at the BS. The Random is the case when all the channel elements, except the $m_\mathrm{Tx}$-th column replaced with $\bh_{\mathrm{SRS}}$, are randomly distributed following $\cC\cN(0,1)$. As a baseline, we compare the conventional scenario with $\bF=\bP\bH_\mathrm{CSI-RS}$. This baseline is denoted as ``Type 1'' in the following figures. We also compare the upper bound of conventional method without quantization loss as $\bF=\bP \bV_\mathrm{CSI-RS-uq}$ where $\bV_\mathrm{CSI-RS-uq}$ is the right singular matrix of $\bH_\mathrm{CSI-RS-uq}$. Since this is the upper bound of the Type 2 codebook, we denote this as ``Type 2'' in the figures.  We also have the ideal case with $ \bF={\bV}_\mathrm{Ideal}$ where ${\bV}_\mathrm{Ideal}$ is the right singular matrix of the true downlink channel $\bH^\mathrm{H}$.
%\begin{table*}
%	\renewcommand{\arraystretch}{2} % 2column=2
%	\centering
%	\caption{Average spectral efficiency [bps/Hz] according to the reconstruction techniques when $\bw_k$ is dynamically selected based on the knowledge of $\bh_{\mathrm{SRS}}$, and $L=M_\mathrm{UE}=4$ layer feedback is assumed. }
%	\begin{tabular}{c| c c c c c c}
%		\multirow{2}*{\shortstack{BS \\ Antenna \\ Structure}}
%		& CSI-RS & Random & Ratio & IP-max & Element & Structure \\
%		&Pre-ULA   & Pre-UPA  & Pinv-ULA  & Pinv-UPA & Ideal & \\
%		\hline\hline
%		\multirow{2}*{ULA} 
%		& 23.9382 & 20.6688 & 25.5891 & 21.0966 & 21.0966 & 17.2927 \\
%		& 27.2052 & 26.9093 & \textbf{27.2342} & 26.9294 & \textbf{31.3884} &\\
%		\hline
%		\multirow{2}{*}{UPA}
%		& 18.4534 & 22.3962 & 20.8487 & 21.6849 & 21.6849 & 22.0801 \\
%		& 26.8392 & \textbf{27.8958} & 26.9011 & 27.8669 & \textbf{34.2036} &\\
%		\hline
%	\end{tabular}
%	
%	\label{ULA,UPA,beamformed average rate}
%\end{table*}

\begin{figure}
	\centering
	\includegraphics[width=0.95\columnwidth]{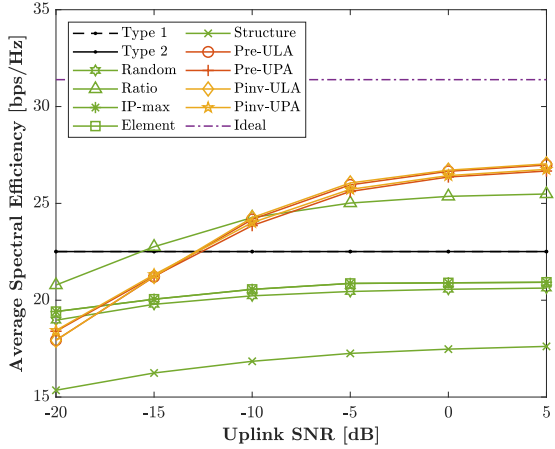} % 0.95 for two col / 0.55 for one col
	\caption{Average spectral efficiency of the different channel reconstruction techniques according to $\rho_\mathrm{UL}$ with the ULA and fixed widebeam at the BS. The $N_\mathrm{BS} \times M_\mathrm{UE}$ downlink channel is reconstructed through $L=M_\mathrm{UE}=4$ layer feedback.}
	\label{ULA-widebeam-all-fig}
\end{figure}

\begin{figure}
	\centering
	\includegraphics[width=0.95\columnwidth]{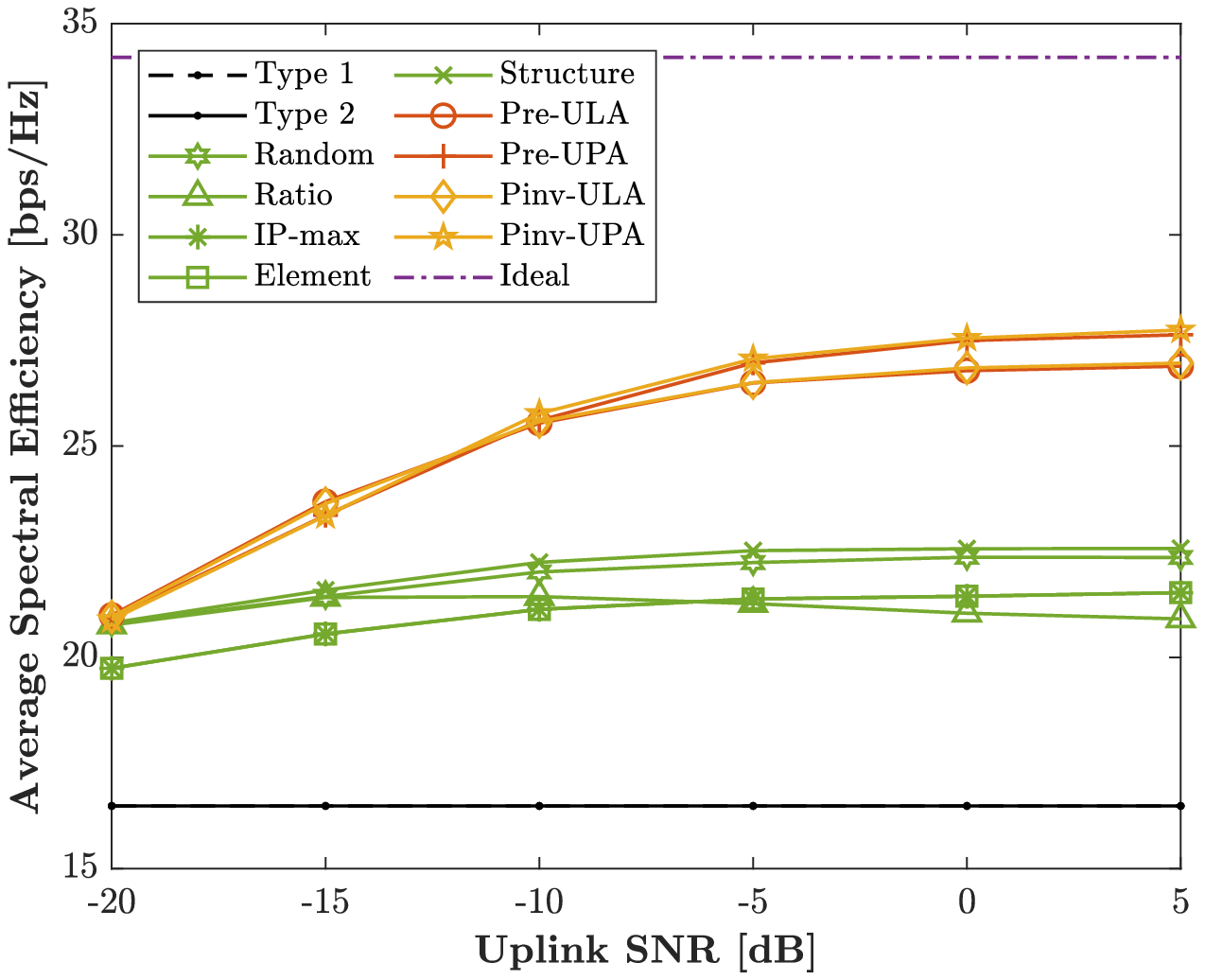} % 0.95 for two col / 0.55 for one col
	\caption{Average spectral efficiency of the different channel reconstruction techniques according to $\rho_\mathrm{UL}$ with the UPA and fixed widebeam at the BS. The $N_\mathrm{BS} \times M_\mathrm{UE}$ downlink channel is reconstructed through $L=M_\mathrm{UE}=4$ layer feedback.}
	\label{UPA-widebeam-all-fig}
\end{figure}

In Figs. \ref{ULA-widebeam-all-fig} and \ref{UPA-widebeam-all-fig}, we consider the case when the BS adopts a fixed widebeam for $\bw_k$ without any prior information of channel. We design the widebeam with boresight $0\degree$ and beamwidth about $40\degree$ as in \cite{Lee:2019SPAWC}. We consider $L=M_{\mathrm{UE}}=4$ for the feedback layer and the $N_\mathrm{BS} \times M_{\mathrm{UE}}$ full MIMO downlink channel reconstruction. Fig. \ref{ULA-widebeam-all-fig} shows the average spectral efficiency of proposed downlink channel reconstruction techniques according to the uplink SNR $\rho_\mathrm{UL}$ assuming the ULA at the BS. It can be observed that the \textit{pre-search technique} and \textit{pseudo-inverse technique} outperform the other channel reconstruction techniques. It is better for these two techniques to assume the ULA for the reconstruction since the BS is deployed with the ULA in this scenario. Despite its simplicity, the \textit{ratio technique} shows quite good performance because of the simple structure of the ULA. As we discussed in Section \ref{Channel Reconstruction Methods}, the \textit{IP maximization}, \textit{element-wise}, and \textit{structure techniques} show poor performance, comparable to the Random case, because of not considering any channel structure or too much degree-of-freedom in the optimizations. Especially, the performance difference between the \textit{structure} and \textit{pre-search techniques} clearly shows that it is essential to have judicious preprocessing before the optimization for downlink channel reconstruction. Note that the Type 1 and Type 2 cases show the same performance since the codewords of PMI codebook are unitary matrices when~$L=M_\mathrm{UE}$.

\begin{figure}
	\centering
	\includegraphics[width=0.95\columnwidth]{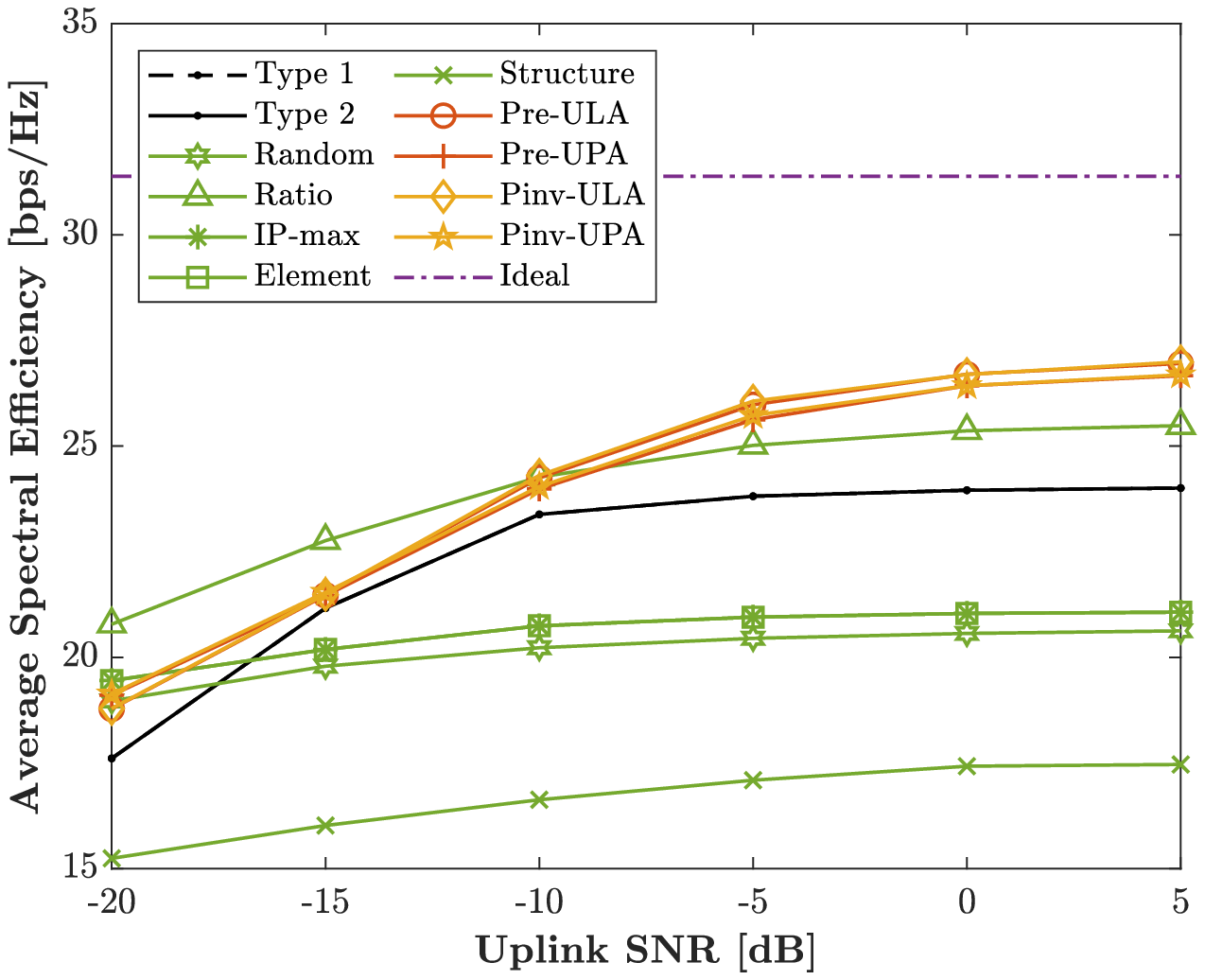} % 0.95 for two col / 0.55 for one col
	\caption{Average spectral efficiency of the different channel reconstruction techniques according to $\rho_\mathrm{UL}$ with the ULA and dynamically selected beam at the BS. The $N_\mathrm{BS} \times M_\mathrm{UE}$ downlink channel is reconstructed through $L=M_\mathrm{UE}=4$ layer feedback.}
	\label{ULA-beamformed-all-fig}
\end{figure}
\begin{figure}
	\centering
	\includegraphics[width=0.95\columnwidth]{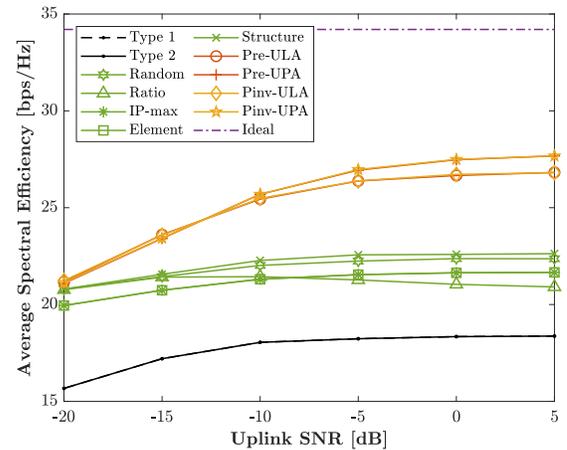} % 0.95 for two col / 0.55 for one col
	\caption{Average spectral efficiency of the different channel reconstruction techniques according to $\rho_\mathrm{UL}$ with the UPA and dynamically selected beam at the BS. The $N_\mathrm{BS} \times M_\mathrm{UE}$ downlink channel is reconstructed through $L=M_\mathrm{UE}=4$ layer feedback.}
	\label{UPA-beamformed-all-fig}
\end{figure}

In Fig. \ref{UPA-widebeam-all-fig}, we considered the UPA at the BS. The \textit{pre-search technique} and \textit{pseudo-inverse technique} still outperform the other channel reconstruction techniques, and the figure shows that it now becomes better for these techniques to assume the UPA for the reconstruction. Note that the \textit{pseudo-inverse technique} is much more practical than the \textit{pre-search technique} since it only requires basic matrix operations. Unlike in Fig. \ref{ULA-widebeam-all-fig}, the \textit{ratio technique} does not perform well since the reconstruction procedure of the \textit{ratio technique} is not suitable to the UPA case. Through Figs. \ref{ULA-widebeam-all-fig} and \ref{UPA-widebeam-all-fig}, it can be observed that the \textit{structure technique} has poor performance in spite of its high complexity since it has no prior information about the dominant AoAs/AoDs and just set those randomly. 
In Figs. \ref{ULA-beamformed-all-fig} and \ref{UPA-beamformed-all-fig}, we consider the case when the BS dynamically selects $\bw_k$ as in (\ref{j_max}). We set $L=M_{\mathrm{UE}}=4$ feedback layer for the $N_\mathrm{BS} \times M_{\mathrm{UE}}$ full MIMO downlink channel reconstruction. Fig. \ref{ULA-beamformed-all-fig} shows the average spectral efficiency of proposed downlink channel reconstruction techniques according to $\rho_\mathrm{UL}$ assuming the ULA at the BS.
Although the CSI-RS beamforming matrix $\bP$, which is a function of $\bw_k$, is now dynamically selected, the figure shows that there is no noticeable difference on the performance of the proposed techniques compared to Fig.~\ref{ULA-widebeam-all-fig} since the data transmission beamformers of the proposed techniques are already adjusted with the reconstructed channel and independent of $\bP$. The Type 1 and Type 2 cases, however, become better than Fig. \ref{ULA-widebeam-all-fig} as $\rho_\mathrm{UL}$ increases. This is because the BS exploits the prior channel knowledge $\bh_{\mathrm{SRS}}$ not only for the CSI-RS beamforming but also for the data transmission. Still, the proposed \textit{ratio}, \textit{pre-search} and \textit{pseudo-inverse techniques} outperform the Type~1 and Type 2 cases.
\begin{figure}
	\centering
	\includegraphics[width=0.95\columnwidth]{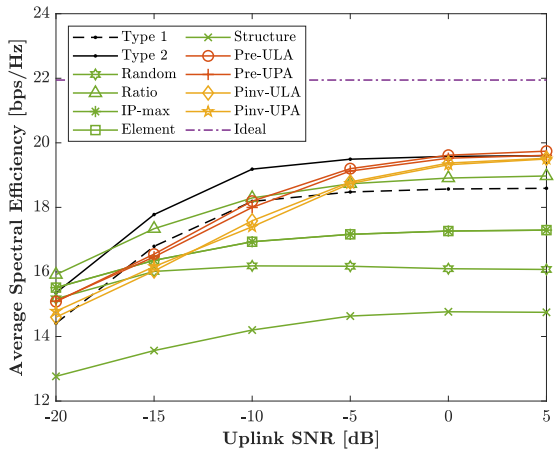} % 0.95 for two col / 0.55 for one col
	\caption{Average spectral efficiency of the different channel reconstruction techniques according to $\rho_\mathrm{UL}$ with the ULA and dynamically selected beam at the BS. The $N_\mathrm{BS} \times L$ downlink channel is reconstructed through $L=2$ layer feedback.}
	\label{ULA-beamformed-all-layer2-fig}
\end{figure}
\begin{figure}
	\centering
	\includegraphics[width=0.95\columnwidth]{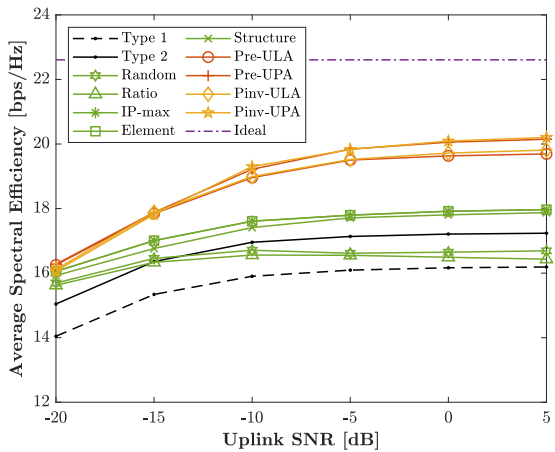} % 0.95 for two col / 0.55 for one col
	\caption{Average spectral efficiency of the different channel reconstruction techniques according to $\rho_\mathrm{UL}$ with the UPA and dynamically selected beam at the BS. The $N_\mathrm{BS} \times L$ downlink channel is reconstructed through $L=2$ layer feedback.}
	\label{UPA-beamformed-all-layer2-fig}
\end{figure} 

In Fig. \ref{UPA-beamformed-all-fig}, we plot the average spectral efficiency of proposed downlink channel reconstruction techniques according to $\rho_\mathrm{UL}$ assuming the UPA at the BS. Similar to Fig. \ref{ULA-beamformed-all-fig}, there is spectral efficiency improvement of the Type 1 and Type 2 cases compared to that in Fig. \ref{UPA-widebeam-all-fig} as $\rho_\mathrm{UL}$ increases; however, the proposed techniques still experience no noticeable difference in terms of their spectral efficiencies.

\begin{figure}
	\centering
	\includegraphics[width=0.95\columnwidth]{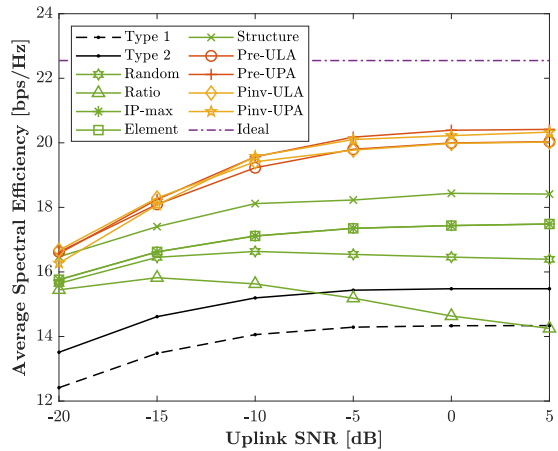} % 0.95 for two col / 0.55 for one col
	\caption{Average spectral efficiency of the different channel reconstruction techniques according to $\rho_\mathrm{UL}$ with the UPA and dynamically selected beam at the BS. The UMa scenario was considered for the SCM channel. The $N_\mathrm{BS} \times L$ downlink channel is reconstructed through $L=2$ layer feedback.}
	\label{UPA-beamformed-all-layer2-UMa-ULSNR-change-fig}
\end{figure}
\begin{figure}
	\centering
	\includegraphics[width=0.95\columnwidth]{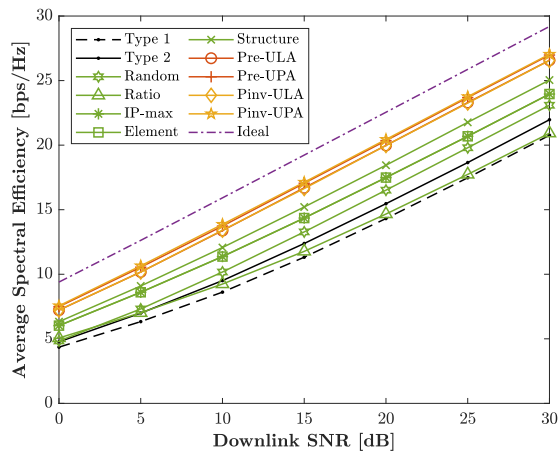} % 0.95 for two col / 0.55 for one col
	\caption{Average spectral efficiency of the different channel reconstruction techniques according to $\rho_\mathrm{DL}$ with the UPA and dynamically selected beam at the BS. The UMa scenario was considered for the SCM channel with the fixed uplink SNR $\rho_\mathrm{UL}=0$ dB. The $N_\mathrm{BS} \times L$ downlink channel is reconstructed through $L=2$ layer feedback.}
	\label{UPA-beamformed-all-layer2-UMa-DLSNR-change-fig}
\end{figure} 

Figs. \ref{ULA-beamformed-all-layer2-fig} and \ref{UPA-beamformed-all-layer2-fig} consider the same scenario as in  Figs. \ref{ULA-beamformed-all-fig} and~\ref{UPA-beamformed-all-fig} except $L=2$ for the PMI feedback layer. The BS then tries to reconstruct the $N_\mathrm{BS} \times L$ MIMO downlink channel. The figures show the average spectral efficiency of the proposed downlink channel reconstruction techniques according to $\rho_\mathrm{UL}$ assuming the ULA/UPA at the BS. Since the BS transmits data only through $L=2$ layer spatial multiplexing, it can be observed that the spectral efficiencies are lower than the previous cases of $L=M_\mathrm{UE}=4$. The overall trends among the proposed downlink channel reconstruction techniques, however, are similar to those of Figs. \ref{ULA-beamformed-all-fig} and \ref{UPA-beamformed-all-fig}. Since the \textit{pre-search technique} and \textit{pseudo-inverse technique} perform quite well with $L=2$, we can conclude that the proposed techniques are able to reconstruct the downlink channel even when $L$ is less than $M_\mathrm{UE}$. Note that, although the Type 2 case does not assume any CSI quantization error, the \textit{pre-search technique} and \textit{pseudo-inverse technique} outperform the Type~2 case when the BS is equipped with the UPA. This clearly shows the loss of conventional methods by only using the low dimensional effective CSI-RS channel for spatial multiplexing.

In Figs.  \ref{UPA-beamformed-all-layer2-UMa-ULSNR-change-fig} and \ref{UPA-beamformed-all-layer2-UMa-DLSNR-change-fig}, we adopt a different scenario of urban macro (UMa) with the same carrier frequency for the SCM channel. In Fig. \ref{UPA-beamformed-all-layer2-UMa-ULSNR-change-fig}, we plot the average spectral efficiency with the same assumptions as in Fig. \ref{UPA-beamformed-all-layer2-fig}. It is clear from the figure that the overall trends among the proposed downlink channel reconstruction techniques are the same with the UMi scenario. In Fig. \ref{UPA-beamformed-all-layer2-UMa-DLSNR-change-fig}, we plot the average spectral efficiency with the downlink SNR $\rho_\mathrm{DL}$ with the fixed uplink SNR $\rho_\mathrm{UL}=0$ dB while other assumptions are the same as in Fig. \ref{UPA-beamformed-all-layer2-UMa-ULSNR-change-fig}. The figure shows the proposed techniques, especially the \textit{pre-search} and \textit{pseudo-inverse techniques}, work well for all range of $\rho_\mathrm{DL}$.

\begin{figure}
	\centering
	\includegraphics[width=0.95\columnwidth]{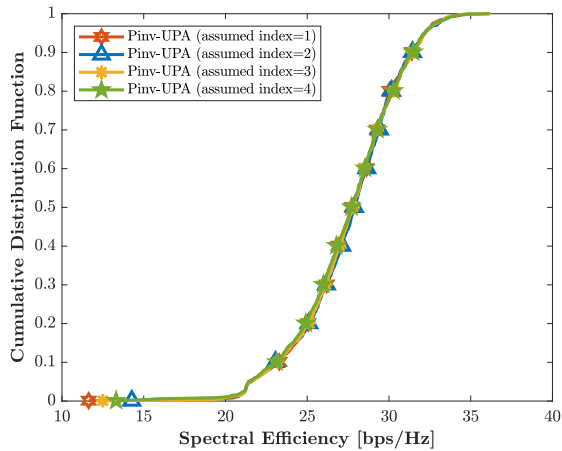} % 0.95 for two col / 0.55 for one col
	\caption{Spectral efficiency CDF of the \textit{pseudo-inverse technique} according to differently assumed $m_\mathrm{Tx}$ while the true value is $m_\mathrm{Tx}=1$. The BS is deployed with the UPA using the fixed widebeam for the CSI-RS beamforming, and $\rho_\mathrm{UL}=0$ dB is assumed. The $N_\mathrm{BS} \times M_\mathrm{UE}$ downlink channel is reconstructed through $L=M_\mathrm{UE}=4$ layer feedback.}
	\label{UPA-widebeam-index-fig}
\end{figure}

In Fig. \ref{UPA-widebeam-index-fig}, we plot the spectral efficiency cumulative distribution function (CDF) of the \textit{pseudo-inverse technique} assuming the UMi scenario and the UPA at the BS to see the effect of imperfect knowledge of $m_\mathrm{Tx}$. We consider the widebeam weight for $\bw_k$ and $L=M_{\mathrm{UE}}=4$ layer feedback as in Figs. \ref{ULA-widebeam-all-fig} and~\ref{UPA-widebeam-all-fig}. We set the true transmit antenna index of the UE $m_\mathrm{Tx}$ as~1 while the BS assumes different values of $m_\mathrm{Tx}$ for the downlink channel reconstruction. It is clear from the figure that the knowledge of $m_\mathrm{Tx}$ does not affect much on the spectral efficiency performance as discussed in Section~\ref{effect of m_Tx}.

\section{Conclusion} \label{conclusion}
In this paper, we proposed possible downlink massive MIMO channel reconstruction techniques at the BS. Considering practical antenna structures to reduce the downlink CSI-RS overhead, the proposed techniques work in TDD by exploiting both the downlink CSI-RS and the uplink SRS. The numerical results showed that the spectral efficiencies by spatial multiplexing based on the proposed downlink channel reconstruction techniques outperformed the conventional methods of using the fed back PMI directly in most cases. Among the proposed techniques, the \textit{pre-search technique} and \textit{pseudo-inverse technique} outperformed the other techniques in terms of the spectral efficiency while the \textit{pseudo-inverse technique} is much more practical due to its low complexity. In addition, we showed that the proposed channel reconstruction techniques are not affected by the imperfect knowledge of the transmit antenna index of the UE at the BS.

Possible future research directions would include practical symbol detection techniques at the UE assuming the BS may not have perfect knowledge of the transmit antenna index of the UE, and downlink channel reconstruction for the case when the UE has multiple transmit antennas. It is also worth investigating the performance limit of downlink channel reconstruction using the CSI-RS and SRS to analyze how close the proposed techniques to the limit.

%\appendices
%% you can choose not to have a title for an appendix
%% if you want by leaving the argument blank
%\section{}

%\section*{Acknowledgment}


\bibliographystyle{IEEEtran}
\bibliography{refs_downlink_TWC}

\end{document}